\documentstyle[twocolumn,pra,aps,epsfig]{revtex}
\begin{document}
\flushbottom

\draft
\title{Generation of arbitrary Dicke states in spinor Bose-Einstein
condensates}
\author{S. Raghavan$^1$, H. Pu$^2$, P. Meystre$^2$ and N.P. Bigelow$^{3}$}
\address{
{$^{1}$Rochester Theory Center for Optical Science
and Engineering} \\
{and Department of Physics and Astronomy,}
{University of Rochester, Rochester, New York 14627}\\
{$^2$Optical Sciences Center, The University of Arizona,
   Tucson, Arizona 85721}\\
{$^3$ Laboratory for Laser Energetics and Department of Physics and
Astronomy} \\
{University of Rochester, Rochester, New York 14627}\\ (\today)
\\ \medskip}\author{\small\parbox{14.2cm}{\small\hspace*{3mm}
We demonstrate that the combination of two-body collisions and
applied Rabi pulses makes it possible to prepare
arbitrary Dicke (spin) states as well as maximally entangled states by
appropriate sequencing of external fields. \\
\\[3pt]PACS numbers: 03.75.Fi, 05.30.Jp }}
\maketitle

\narrowtext
It is now widely recognized that the use of squeezed atomic states
has the potential for substantial improvement in the sensitivity
of atom interferometers. This realization has led to much
theoretical and experimental work on schemes to realize atomic spin
squeezing. Building upon the seminal work on photon number
squeezing by Kitagawa and Yamamoto \cite{kitayama86}, Kitagawa and
Ueda proposed and clarified the basic issues concerning the
definition and preparation of spin squeezed states
(SSS's)\cite{kitaueda91,kitaueda93}. This was followed by studies
of the production of squeezed atomic states
\cite{wineland92,wineland94}, the transfer of squeezing from
incident squeezed light to an assembly of cold atoms
\cite{kuzprl97,hald99}, and the measurement of a collection of
squeezed atoms \cite{sorensen97,saito97,kuzepl98,kuzpra99,taka99}.
So far, though, this work seems limited to the quantum state
control of just a few atoms or ions.

At the same time, Bose-Einstein condensates have become almost
routinely available in the laboratory. They provide us with large
coherent ensembles of ultracold atoms which seem ideal to
perform quantum state manipulation experiments from an exquisitely
well-controlled initial state. In particular, recent experiments
on two-component Bose-Einstein condensates (BECs) in $^{87}$Rb
\cite{hall98-2} have led to considerable work on the dynamics of
the relative phase and number fluctuations of Bose condensates.
There have been vigorous efforts on the issue of temporal phase
coherence between two coupled Bose condensates. A number of
workers have thought of this system as a prototypical Josephson
junction between two weakly-coupled superfluid systems.

While much of the theoretical work so far amounts to what are
essentially semiclassical (mean-field) analysis
\cite{milburn97,smerzi97,raghavan99,williams99} and studies
involving small quantum corrections going beyond the mean-field
Gross-Pitaevskii equation
\cite{milburn97,villain97,rag-smer-ken99,java99,smerrag2000,steel98},
some publications have gone past this framework. For instance,
Villain {\em et al.} \cite{villain97}, Steel and Collett
\cite{steel98}, and Javanainen and Ivanov \cite{java99} have
studied dynamical aspects of the quantum state of a coupled BEC.
There has also been intensive work on spin manipulation in the
context of spinor BECs, motivated by the experimental work on
$^{23}$Na by the group at MIT \cite{stamperstenger98,miesner99}.
Theoretical analyses, notably that of Law {\em et al.} have
addressed the issue of collective spin properties in spinor BECs
\cite{lawpubig98}. Recently, S{\o}rensen {\em et al.} studied spin
squeezing in a spinor condensate due to the internal nonlinear
atom-atom interaction\cite{zzz}. In this note, we extend these
results by describing a quantum control technique which allows to
prepare {\em arbitrary} Dicke spin states, as well as maximally
entangled states by means of properly sequencing of external
coupling fields.

To set the stage for the discussion and relate it to the physics
of Bose-Einstein condensates, we first recall some basic
properties of coherent spin states (CSS's).\cite{radcliffe71acgt72}
Consider a two-component condensate with $N_i$ atoms in component
$i$, and $N_1+N_2=N$.  A general state of the system can be
written as a superposition of number difference states
\begin{equation} \label{eq:state}
|{\Psi} \rangle = \sum_{N_2=0}^{N} c_{N_2} |{N_1,N_2} \rangle.
\end{equation}
Alternatively, we may introduce the effective angular momentum
quantum number $j=(N_1+N_2)/2=N/2$ and the $z$-projection quantum
number $m= (N_2 - N_1)/2$ and reexpress the state $|{\Psi}
\rangle$ in terms of angular momentum states $|j,\,m \rangle$. In
particular, we have the correspondence
\begin{equation}
|j,-j\rangle = |N_1=N,N_2=0 \rangle
\end{equation}
a state which can be thought of as the ``ground state'' of
coherent spin states. Our goal is to construct arbitrary states of the
system from this ground state $|j,-j\rangle$.

In a landmark paper, Kitagawa and Yamamoto \cite{kitayama86}
showed that photons could be number-squeezed by making use of a
Kerr nonlinear medium. Their procedure consisted of splitting a
coherent beam, passing one of its components through a Kerr
nonlinear medium, and interfering the resulting light with the
other coherent beam in a Mach-Zehnder interferometer. The enhanced
squeezing achieved in that case results from the fact that it
involves a quartic four-wave process, while ordinary squeezing is
a quadratic process. Following this work, Kitagawa and Ueda
\cite{kitaueda91,kitaueda93} showed that it was possible to
likewise produce squeezed spin states (SSS's) by making use of spin
Hamiltonians quadratic in the spin operators. Our proposed scheme
relies similarly on the existence of such a term in the
Hamiltonian describing the dynamics of a coupled two-component
condensate.

We consider for simplicity the so-called two-mode model describing
the coupled two-component condensate. This approximation consists
in neglecting all modes except the condensate modes. Its second
quantized Hamiltonian is
\begin{eqnarray} \label{eq:ham}
H &=& E_1 b_1^{\dagger}b_1 +E_2 b_2^{\dagger}b_2 +\frac{u_{11}}{2}
(b_1^\dagger b_1^\dagger b_1 b_1) + \frac{u_{22}}{2} (b_2^\dagger
b_2^\dagger b_2 b_2) \nonumber \\ && + u_{12} (b_1^\dagger b_1
b_2^\dagger b_2) - \frac{1}{2}({\Omega} b_1 b_2^\dagger +
{\Omega}^* b_2^\dagger b_1),
\end{eqnarray}
where $b_i$ is the annihilation operator for component $i$,
\begin{mathletters}
\label{eq:defn}
\begin{equation}
E_i = \int d^3 r\,\phi_i^*({\bf r})\left[\hat{{\bf
p}}^2/(2m)+V_i({\bf r}) \right] \phi_i({\bf r})\
\end{equation}
is the single-particle energy of mode $i$, $V_i ({\bf r})$ being
the trapping potential, and
\begin{equation}
u_{ij} = \frac{4\pi \hbar^2 a_{ij}}{m} \int d^3 r \,|{\phi}_i({\bf
r})|^2 |{\phi}_j({\bf r})|^2\,
\end{equation}
\end{mathletters}
describes the two-body collisions in the condensate in the
$s$-wave scattering approximation. Here $a_{ij}$ is the scattering
length for a two-body collision between an atom of $i^{\rm{th}}$
component and that of $j^{\rm{th}}$ component and $\phi_i({\bf
r})$ represents the condensate wave function for mode $i$. Finally
\begin{equation}
{\Omega} = \Omega_0 \int d^3  r \, {\phi}_1({\bf r})
{\phi}_2^*({\bf r})
\end{equation}
describes the strength of the linear coupling between components.
The procedure to obtain the Hamiltonian (\ref{eq:ham}) and the
discussion of the validity of the two-mode approximation can be
found for instance in Refs.
\cite{milburn97,villain97,steel98,gordon99}. We note that this
same Hamiltonian has been previously considered by Scott {\em et
al.} \cite{scott86bernstein90} in quantum studies of
self-localization.

The analysis of Eq.~(\ref{eq:ham}) is greatly simplified by the
introduction of the angular momentum operators
\begin{mathletters}
\begin{eqnarray}
\hat{J}_x &=& \frac{1}{2}(b_1^\dagger b_2 + b_2^\dagger b_1)\, ,\\
\hat{J}_y &=& \frac{i}{2}(b_1^\dagger b_2 -b_2^\dagger b_1)\, ,\\
\hat{J}_z &=& \frac{1}{2}(b_2^\dagger b_2 - b_1^\dagger b_1)\, ,
\end{eqnarray}
\end{mathletters}
in terms of which Hamiltonian (\ref{eq:ham}) may be rewritten as
\begin{equation} \label{eq:ham1}
H = {\kappa} \hat{J}_z^2 -{\Omega}_x \hat{J}_x -{\Omega}_y
\hat{J}_y,
\end{equation}
where we have introduced the effective nonlinear coupling
\begin{equation} \label{eq:ustyle}
{\kappa} = \frac{1}{2}(u_{11} + u_{22}) - u_{12},
\end{equation}
and the real Rabi couplings ${\Omega}_x = Re({\Omega}),{\Omega}_y
= Im({\Omega})$. Note that some control of the nonlinear parameter
$\kappa$ can be achieved through the proper engineering of the
trapping potential, and hence of the condensate wave functions
$\phi_i({\bf r})$. Another way to adjust $\kappa$ is through tuning
the scattering lengths via Feshbach resonances, although this is not
always easy to do in practice. However, the purpose of the present
work is to show that spin squeezing can be controlled and manipulated
with the external coupling fields instead of the internal nonlinearity
of the condensate.
In deriving (\ref{eq:ham1}), we have assumed that
\[ E_1-E_2+(N-1/2)(u_{11}-u_{22})=0\]
a condition that can always be achieved by shifting the energy
levels of the condensate components. If this condition is not
fulfilled the Hamiltonian (\ref{eq:ham1}) contains an additional
term proportional to $\hat{J}_z$.

Importantly for the following discussion, we observe that the
independent temporal control of the Rabi pulses characterized by
${\Omega}_x$ and ${\Omega}_y$ is a well-established experimental
tool. Depending on the particular system of interest, these pulses
can be in the form of laser light, radio-frequency microwaves or
magnetic fields.

With the Hamiltonian (\ref{eq:ham1}) at hand, we now demonstrate
how an appropriate choice of the external coupling fields allows
one to generate arbitrary Dicke states from the ground state
$|j,-j\rangle$. We first observe that a general CSS
$|\theta,\varphi \rangle$ can be created by rotating
$|j,-j\rangle$ by the angle $\theta$ about the axis $\vec{n}=(\sin
\varphi,\,-\cos \varphi,\,0)$,
\begin{eqnarray} \label{eq:cohstate}
|{\theta,\varphi} \rangle &=& \exp [-i \theta(\hat{J}_x
\sin\varphi - \hat{J}_y \cos \varphi)]\, |{j,-j} \rangle
\end{eqnarray}
Starting from state $|j,-j \rangle$, switching on the coupling
pulses and properly adjusting the strength of ${\Omega}_x$ and
${\Omega}_y$, we can prepare any CSS. For the CSS $|\theta,
\varphi \rangle$, the expectation value and the variance of
operator $\hat{J}_z$ are
\begin{eqnarray*}
\langle \hat{J}_z \rangle &=& -j \cos \theta \, ,\\
\langle \Delta \hat{J}_z^2 \rangle &=& \frac{j}{2}\sin^2 \theta\,.
\end{eqnarray*}
(Note that neither $\langle \hat{J}_z \rangle$ nor $\langle \Delta
\hat{J}_z^2 \rangle$ will change after the coupling fields are
turned off.)  Following the usual convention we say that a
$j$-spin state is {\em squeezed along} $\hat{J}_z$ if the state has
the same $\langle \hat{J}_z \rangle $ but reduced $\langle \Delta
\hat{J}_z^2 \rangle$ compared to the CSS $|\theta, \varphi
\rangle$. Or we can define a parameter \cite{foot0}
\begin{equation}
\label{parameter}
\xi_z = \frac{2j \langle \Delta \hat{J}_z^2 \rangle }
{j^2- \langle \hat{J}_z \rangle^2}\,.
\end{equation}
A state is squeezed along $\hat{J}_z$ if $\xi_z < 1$.

Procedures to achieve spin squeezing in the equatorial plane
$\langle {\hat J}_z\rangle = 0$ was previously proposed by
Kitagawa and Ueda\cite{kitaueda91,kitaueda93}, and
more recently by Law {\em et al.} \cite{law00}. We briefly
recall them as a preparation for our general scheme. In the first
scheme, a pulse is applied on the ground state $|j,-j\rangle$
along the $y$-direction at $t=0$. The duration of the pulse is
assumed to be short enough that the effects of the nonlinear
interaction are negligible during this time. This pulse aligns
the spin vector along the negative $x$-axis, with the resultant
state, $|J,-J \rangle_x$. After the pulse, the system evolves
under the Hamiltonian $H_{spin}={\kappa} \hat{J}_z^2$ until a
second short pulse is applied along $x$ at $t=t_1$. At the end of
this pulse, we have
\begin{mathletters}
\begin{eqnarray}
\langle \hat{J}_z \rangle &=& 0\\ \langle \Delta \hat{J}_z^2
\rangle &=& \frac{j}{2}\left\{1 +
\left(\frac{j}{2}-\frac{1}{4}\right) \right. \nonumber \\ 
&& \left. \left[{\cal A} - \sqrt{{\cal
A}^2 + {\cal B}^2} \cos2(\alpha + \delta)\right]\right\},
\label{ueda}
\end{eqnarray}
\end{mathletters}
where $\alpha =  -\int {\Omega}_x dt/\hbar$ is the strength of the
second pulse and $\delta = \frac{1}{2}\tan^{-1}({\cal B}/{\cal
A})$ with
\begin{eqnarray*}
{\cal A} &=& 1 - \cos^{2(j-1)}(2{\kappa} t_1), \label{eq:Adef}\\
{\cal B} &=& 4 \sin(2\alpha) \cos^{2(j-1)}({\kappa} t_1) \sin ({\kappa}
t_1).
\label{eq:Bdef}
\end{eqnarray*}
Optimal squeezing in $\hat{J}_z$ is achieved for $\alpha =
-\delta$ and $|{\kappa}| t_1 =(3/8)^{1/6} j^{-2/3}$, which leads
to $\langle \Delta \hat{J}_z^2 \rangle_{min} \approx
\frac{1}{2}\left(\frac{j}{3}\right)^{1/3}$.

\begin{figure}
\begin{center}
    \includegraphics*[width=0.95\columnwidth,
height=0.55\columnwidth]{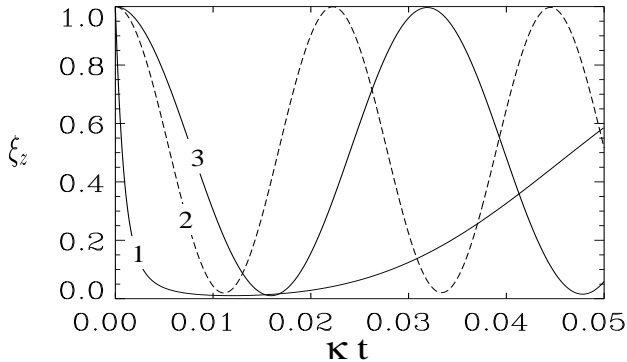}
\vspace{3 mm}
\caption{Time evolution of the parameter $\xi_z$ for $j=500$. Curve 1
corresponds to Eq.~(\ref{ueda}) with $\alpha=-\delta$; the other
two curves are obtained  using the scheme of Ref. [30] . For
curve 2, $\Omega_x/\kappa=20$; for curve 3, $\Omega_x/\kappa=10$.}
\label{fig1}
\end{center}
\end{figure}

As pointed out by Law {\em et al.} Ref.\cite{law00}, an
alternative approach can be realized by switching on a coupling
field along the $x$-direction immediately after the first $\pi/2$
pulse along the $y$-direction, so that the system evolves under
the Hamiltonian $H={\kappa} \hat{J}_z^2 - {\Omega}_x \hat{J}_x$.
During the evolution following the control pulses, $\langle
\hat{J}_z \rangle $ remains equal to 0 while $\langle \Delta
\hat{J}_z^2 \rangle$ can be significantly reduced at appropriate
time. Fig.~\ref{fig1} shows the time evolution of the parameter
$\xi_z$ for both pulse
sequences.

The essential point of both schemes is as already mentioned that
$\langle \hat{J}_z \rangle $ remains equal to zero after the
first $\pi/2$ pulse. This, however, needs not to be the case:  If the
first pulse creates a CSS with $\langle \hat{J}_z \rangle =m_0\neq
0$, then following the above procedures one might obtain a state
squeezed along $\hat{J}_z$ but with  $\langle \hat{J}_z \rangle
\neq m_0$. In order to create squeezed states with arbitrary
prescribed value of $\langle \hat{J}_z \rangle =m_0$ and arbitrary
Dicke states, we propose to use instead the following method:
First, a coupling pulse of appropriate strength is applied to the
ground state $|j,-j \rangle$, creating an initial CSS with
$\langle \hat{J}_z \rangle =m_0$. Assume without loss of
generality that that pulse is along the $x$-axis. Following this
pulse, we then apply a coupling field along the negative $x$-axis.
During the subsequent time evolution, both $\langle \hat{J}_z
\rangle $ and $\langle \Delta \hat{J}_z^2 \rangle$ start to
oscillate. But at certain times when $\langle \hat{J}_z \rangle $
returns back to its initial value $m_0$, $\langle \Delta
\hat{J}_z^2 \rangle$ comes close to a local minimum which is less
than the initial variance, i.e., $\xi_z<1$.
\begin{figure}
\begin{center}
    \includegraphics*[width=0.95\columnwidth,
height=0.55\columnwidth]{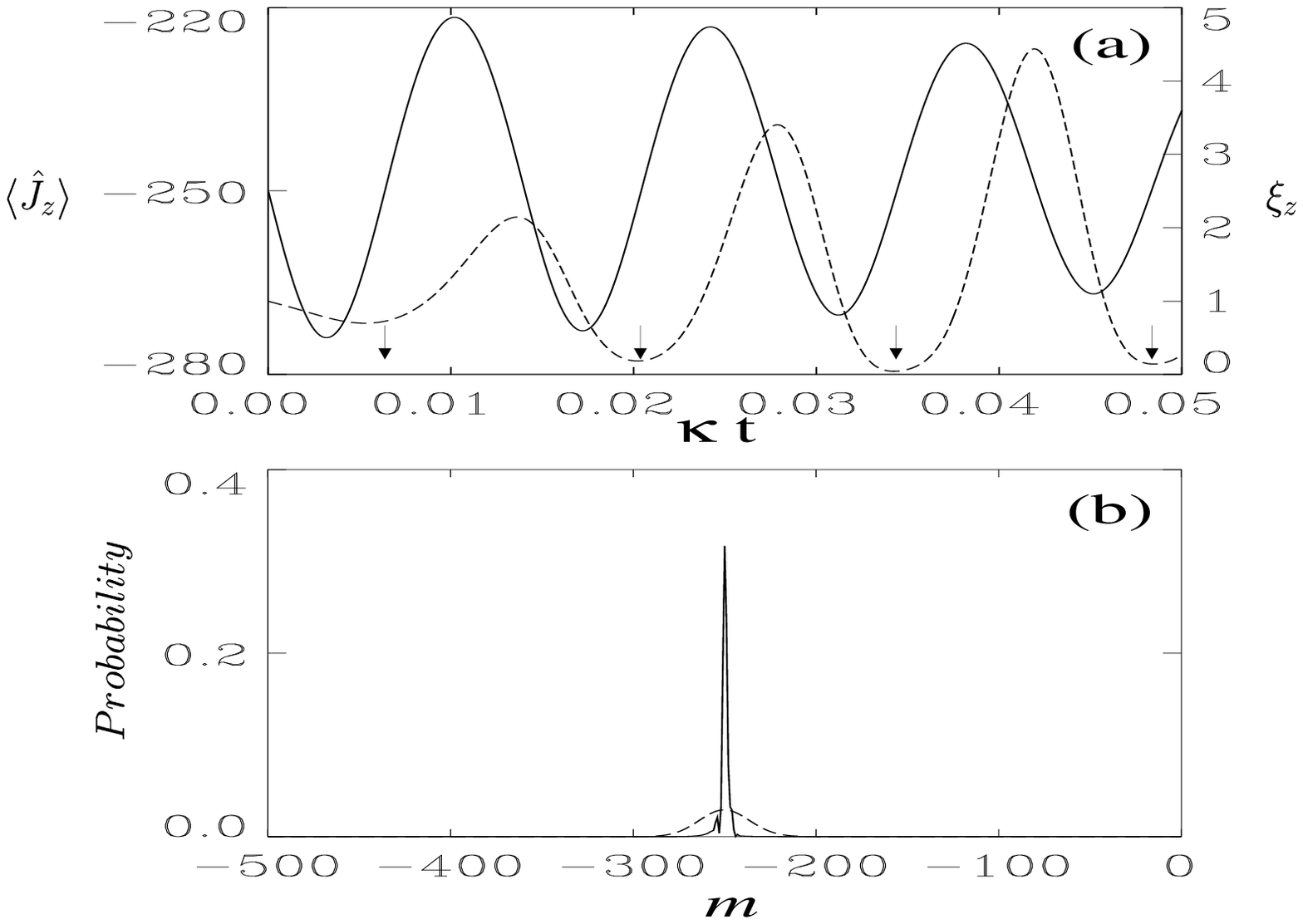}
    \includegraphics*[width=0.95\columnwidth,
height=0.55\columnwidth]{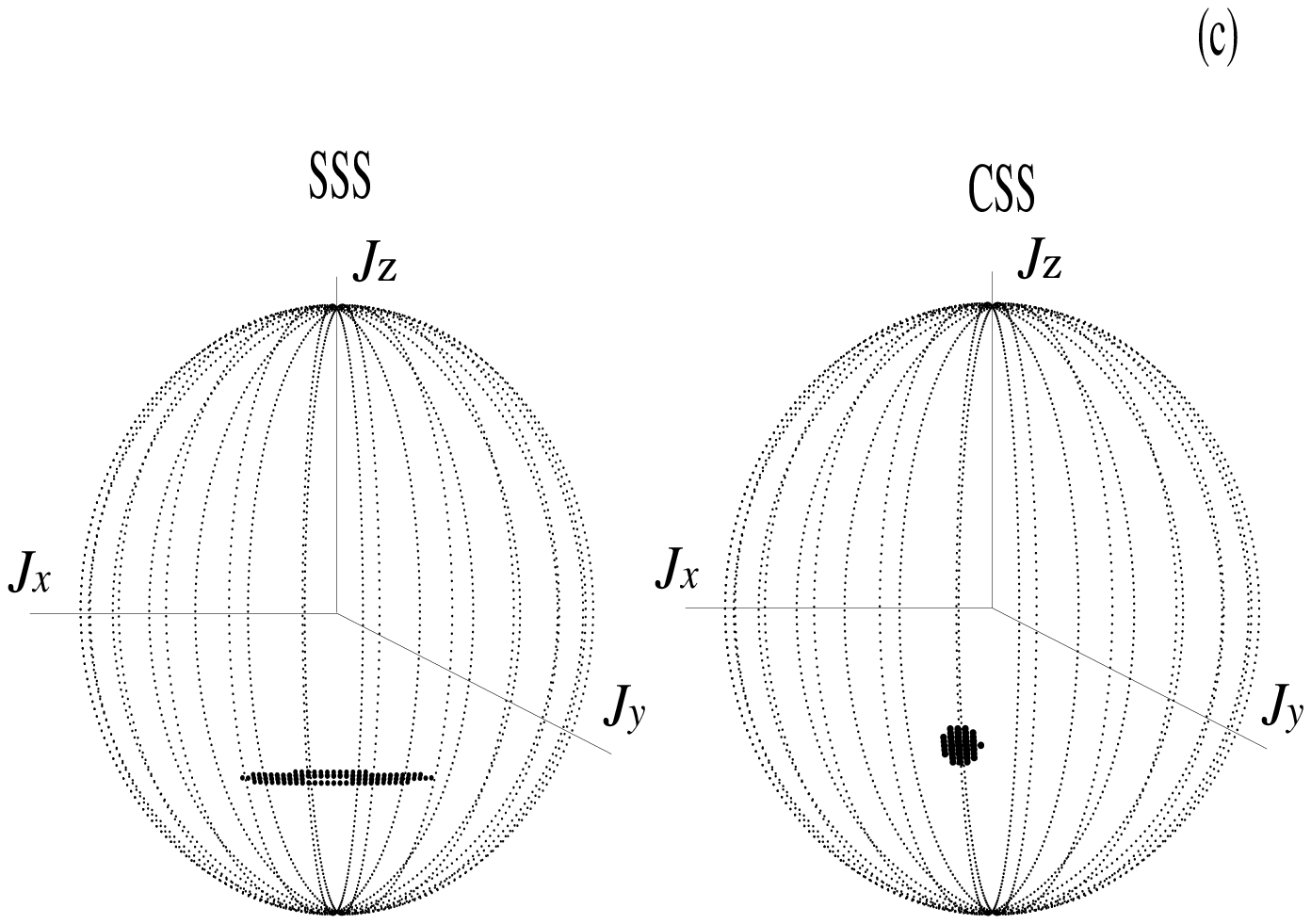}
\vspace{3 mm}
\caption{(a) Time evolution of $\langle \hat{J}_z \rangle$ (left
panel) and $\xi_z$ (right panel).
The initial CSS is $|\theta=\pi/3,\varphi=0 \rangle$ with
$j=500$. After the initial CSS is created, the system evolves
under the Hamiltonian $H=\kappa \hat{J}_z^2 -\Omega_x \hat{J}_x$
with $\Omega_x/\kappa =-30$. At times indicated by the arrows,
$\langle \hat{J}_z \rangle$ reaches the initial value (-250) while
$\xi_z$ is reduced.
(b) Probability distribution of the azimuthal number $m$. Dashed
curve: the initial CSS $|\theta=\pi/3,\varphi=0 \rangle$; solid
curve: the spin squeezed state at $\kappa t=0.0344$.
(c) Quasiprobability distribution on the Bloch sphere for the initial
CSS and the squeezed state at $\kappa t=0.0344$.}
\label{fig2}
\end{center}
\end{figure}
Provided that the coupling field is
turned off at these precise times, a state squeezed along
$\hat{J}_z$ with $\langle \hat{J}_z \rangle =m_0$
will be created. For large enough squeezing, the state
thus prepared can be regarded as an approximation of the Dicke
state $|j,m_0\rangle$ which is the spin analog of
the Fock state. A typical time evolution of
$\langle \hat{J}_z \rangle $ and $\xi_z$ is
shown in Fig.~\ref{fig2}(a) where the initial CSS has $j=500$ and
$\langle \hat{J}_z \rangle =-250$. In this example, the optimal
squeezing occurs at $\kappa t=0.0344$ where the variance
$\langle \Delta \hat{J}_z^2 \rangle$ is reduced by a factor of more than
20. Fig.~\ref{fig2}(b) shows the probability distribution against the
azimuthal number $m$ at the optimal squeezing, while in
Fig.~\ref{fig2}(c) we plot the quasiprobability distribution (QPD)
on a Bloch sphere.

Squeezing and entanglement are oftentimes closely related. In
particular, it is well known that the quadratic Hamiltonian
$H_{spin}= {\kappa} \hat{J}_z^2$ also produces entanglement in a
collective spin system. In particular, starting from the ground
state $|j,-j\rangle_x$, $H_{spin}$ generates at time $t^*=\hbar
\pi/(2{\kappa})$ the state
\begin{eqnarray}
|\Psi(t^*) \rangle &=&
\exp \left(-i \frac{\pi}{2} \hat{J}_z^2 \right) \,|j,-j\rangle_x
\nonumber \\
&=& \frac{1}{\sqrt{2}} \left[ e^{-i\pi/4}|j,-j\rangle_x + (-1)^j
e^{i\pi/4} |j,+j\rangle_x \right].
\label{state}
\end{eqnarray}
The state is called {\em maximally entangled} because if one
individual spin is found to be aligned along the negative $x$-axis
or the positive $x$-axis, so are all other spins. However, the
degree of entanglement degrades after $t^*$, a consequence of the
fact that the state (\ref{state}) is not an eigenstate of the
Hamiltonian $H_{spin}$. To preserve the entanglement
in our system, a $\pi/2$ pulse along the $y$-direction can
be applied
at $t=t^*$. This converts the state $|j,\pm j\rangle_x$
into $|j,\pm j \rangle$. The
resultant state, still maximally entangled, it now an eigenstate
of $H_{spin}$, hence it remains maximally entangled during the
subsequent time evolution. This state is also a Schr\"{o}dinger
cat state since it is a coherent superposition of two macroscopically
distinct states --- one state has all the population in component 1
and the other in component 2.

Maximally entangled states, in particular those of massive
particles instead of fast-escaping photons, are of great
importance in fundamental physics as well as in applications in
quantum information and quantum measurement. A great deal of
effort has been directed toward the creation of such states.
Two-\cite{two1,two2}, three-\cite{arno} and
four-particle\cite{cass} entanglement have been successfully
demonstrated experimentally in trapped ions, Rydberg atoms, and
cavity QED. However, a further increase of the number of entangled
particles in these systems are expected to be a severe
experimental challenge. The two-component, atomic BEC with its
built-in intrinsic nonlinearity appears to be a promising
candidate to generate entanglement on a macroscopic level.

To conclude, we have shown that an arbitrary collective spin
squeezing and entanglement of a two-component spinor condensate
can be readily controlled by the coupling fields between the two
components. Squeezed or entangled spin states will find
applications in high-precision spectroscopy, atomic interferometry
and quantum information, and spinor condensates are attractive
candidates to create such states. In practice, the scheme
presented in this work can be realized in the two-component
$^{87}$Rb condensate\cite{hall98-2}. Another possibility is
offered by the $F=1$ spinor condensate of
$^{23}$Na\cite{stamperstenger98,miesner99} which can be reduced to
an effective spin-(1/2) system by shifting the energy level of the
$m_F=0$ state, e.g., with an ac Stark shift.

\acknowledgments
This work is supported by NSF, the David and Lucile Packard Foundation
and the office of Naval Research.
We thank D.~J.~Heinzen, V.~V.~Kozlov, C.~K.~Law, and
M.~Kitagawa for helpful comments and discussion.  We thank
P. Zoller for making us aware of the related work Ref. [26].


\begin{thebibliography}{10}
\bibitem{foot0} We note that this definition of $\xi_z$ differs from
those used previously [2-5]
We prefer it because of its utility in evaulating not only the spin
squeezed states discussed in [2-5] but for evaluating the Dicke
states discussed here.
\end{thebibliography}

\end{document}